\documentstyle[prb,aps]{revtex}
\begin{document}
\draft
\title{
Electron--lattice kinetics of metals
heated by ultrashort laser pulses}
\author{L. A. Falkovsky and E. G. Mishchenko}
\address{
Landau Institute for Theoretical Physics, Russian Academy of Sciences\\
Kosygina 2, Moscow 117 334, Russia}

\maketitle
\begin{abstract}
We propose a kinetic model of transient nonequilibrium phenomena
in metals
exposed to ultrashort  laser pulses when
heated electrons affect the lattice through  direct
electron--phonon interaction.
This model describes the destruction of a metal under
intense laser pumping.   We derive
the system of equations for the metal, which consists
of hot electrons and a cold lattice.
Hot electrons are described with the help of the Boltzmann equation
and  equation of thermoconductivity.
We use the equations of motion for lattice displacements with
the electron force
included. The lattice deformation is estimated
immediately after the laser pulse
up to the time of electron temperature relaxation.
An estimate shows that the ablation regime
can be achieved.

\end{abstract}
\section{Introduction}
The first theoretical prediction of transient
laser-induced nonequilibrium electron temperature phenomena
in metals was made more than twenty years ago \cite{ABE}.
It was shown that an ultrashort laser pulse ($\sim 10^{-13}-10^{-12}$ s)
produces a nonequilibrium state of the electron gas near a metal surface.
However, experimental picosecond ($\sim 10^{-12}$ s) laser studies of
thermally assisted multiphonon photoemission were unable
to measure, and even failed to observe,
this nonequilibrium electron state \cite{YLB}. This failure had a simple
explanation in terms of the theory of electron--lattice thermal relaxation,
\cite{KLT} which yields a  relaxation time of the
order of $\tau_{e-l}\sim 10^{-12}$ s.
It was necessary to use power pulses shorter than
$\tau_{e-l}$. Such measurements with subpicosecond
($\sim 10^{-13}$ s) pulses revealed a transient
nonequilibrium regime in
transmittivity and IR reflection \cite{FLI} --\cite{ABF},
giant electron emission \cite{KKH} --\cite{KKH'} and the emission of light
\cite{ABG} --\cite{AAM'}.

\par We briefly summarize the physical process.
The ultrashort laser pulse ($\Delta t\sim 10^{-14}-10^{-13}$ s)
absorbed in a metal
raises the electron temperature $T_e$ considerably higher than the lattice
temperature because of the difference
in their specific heats ($c_e \ll c_{\ell}$).
Subsequent electron cooling results mainly from two processes, namely
electron--lattice thermal relaxation and electron thermoconductivity.
These are usually modeled with a set of coupled thermoconductivity
equations for the electron and lattice components.
These equations are nonlinear and can generally  be solved numerically,
yielding the electron temperature relaxation.
The solution also shows  that the
subsequent ablation regime can be achieved,  which involves  the "cold"
destruction of a metal into the parts consisting of different phases.
 "Hot" destruction, namely melting, can also be studied with the help
of this solution \cite{AR}.
However, such an approach has several shorcomings.
First, the question remains as to
whether the equations of thermoconductivity are still hold at such high
frequencies
$(\sim 1/\Delta t)$. Second and more importantly, these equations
 can only describe the
temperature dynamics of a metal but not  electron transport,
lattice deformation, thermionic emission, etc. It is evident that
a strict kinetic approach is needed to describe  the various transport
phenomena properly and derive thoroughly the equation of thermoconductivity
\cite{FM}.

\par In this paper we present a theory of transient nonequilibrium
phenomena in metals
subject to ultrashort laser pulses. Our theory is based on the Boltzmann
equation
for the nonequilibrium electronic partition function. We focus mainly on
times shorter than the electron--lattice relaxation time $\tau_{e-l}$.
Electrons therefore affect the lattice via  direct electron--phonon
interactions. To consider lattice deformations, we use the equations of the so-called
dynamical theory of elasticity. Lattice deformation is due to
 the nonequilibrium electron state and results from
the effective "driving" force (proportional to
$\nabla T_e^2$)  on the lattice.
This force also governs the renormalization (depending on $T_e$) of the lattice
constants (sound velocity and optical phonon gap).
We show that the driving force leads to large lattice deformations,
and can distroy the  crystal. These results are in agreement with
measurements of time-resolved
 X-ray diffraction synchronized with laser pumping \cite{R}.
A nonstationary
increase  in lattice parameters
of Au(111) and Pt(111) single crystals was detected.
Measurements of the shift
 and intensity variation of
 Bragg peaks, as well as  the Debye--Waller factor,
anables one to separate
the effects of lattice deformation and heating. The transformation
of elastic  into  plastic deformation was also observed.

\par The plan of the paper is as follows. In Sec. 2 we present the kinetic
theory of the process under study:
 the Boltzmann equation for an electron gas and the elastic
equation for the lattice are derived, along with the
equation for thermoconductivity.
In Sec. 3, the solutions of the proposed equations
are found for the times of interest. The lattice deformation is
calculated.
In Sec. 4 the solutions  are analyzed.
The lattice deformation is estimated analytically in various limiting cases.
The possibility of  crystal destruction under laser pumping in discussed.

\section{Theoretical Framework}
Let us briefly recapitulate the main equations of our
problem. For the lattice deformation we use the so-called equation of
dynamical theory of elasticity \cite{FM'},\cite{K}
\begin{equation}
\label{elas}
\rho \frac{\partial^2 u_i}{\partial t^2}
-\lambda_{ijlm}\frac{\partial^2 u_l}{\partial x_j \partial x_m}= G_i,
\end{equation}
where $\rho$ is the lattice density, $\lambda_{ijlm}$ is the tensor of elastic
constants, and the driving force describes the effect of free carriers on the
lattice
\begin{equation}
\label{force1}
G_i = \frac{\partial}{\partial x_j} \int\frac{2d^3 p}{(2\pi)^3}
\lambda_{ij}({\bf p}) f_p({\bf r},t).
\end{equation}
The deformation potential $\lambda_{ij}({\bf p})$
yields the change in the local
electron spectrum,
$$\delta \varepsilon ({\bf p,r},t)=\lambda _{ij}({\bf p})u_{ij}
({\bf r},t).$$

To find the electron distribution function $f_p({\bf r},t)$, we
use the Boltzmann equation with the electron--phonon collision integral
\begin{eqnarray}
\label{coll}
{\rm St}f_p=
\sum_n\int \frac{d^3 k}{(2\pi)^3} w^{(n)}_{pk}\delta(\varepsilon_{p+k}-
\varepsilon_p+\omega_k^{(n)})\left((1-f_p)f_{p'}N_{k}^{(n)} -
f_p(1-f_{p'})(1+N_{k}^{(n)})\right)\nonumber\\
+ \sum_n\int \frac{d^3 k}{(2\pi)^3} w^{(n)}_{pk}\delta(\varepsilon_{p+k}-
\varepsilon_p-\omega_k^{(n)})\left((1-f_p)f_{p'}(1+N_{k}^{(n)})-
f_p(1-f_{p'})N_{k}^{(n)}\right),
\end{eqnarray}
with the probability of a scattering process involving a phonon of
the $n$-th branch
$$w_{pk}^{(n)}=\frac \pi {\rho \omega _k^{(n)}}\left| e_i^{(n)}\lambda _{ij}
({\bf p})k_j\right| ^2,$$
where $e_i^{(n)}$ and $\omega_k^{(n)}$  are the polarization and spectrum of
phonons of the $n$-th branch, respectively.
\par Since the  phonon--phonon relaxation time is large ($\sim 10^{-11}$~s)
compared with the times of interest,
the phonon distribution
function $N_k^{(n)}$ takes its equilibrium value
at the lattice temperature $T_{\ell}$,
$$
N_k^{(n)}(T_{\ell})=\frac{1}{\mbox{exp} (\omega^{(n)}/T_{\ell})-1}.
$$
The  electron--electron relaxation time due to  scattering
on phonons $\tau \sim T_{\ell}^{-1} \sim
10^{-14}$~s (see below) is much less than the characteristic time of laser pumping.
Therefore the electron gas is nearly in  thermal
equilibrium at the local temperature $T_e ({\bf r},t) $. We seek
 a solution of the Boltzmann equation in the form
\begin{equation}
\label{loc}
f_p=f_0\left(\frac{\varepsilon_p-\mu}{T_e}\right)+\chi_p\frac{\partial f_0}
{\partial \varepsilon},
\end{equation}
where $f_0$ is the local equilibrium Fermi--Dirac partition function
and $\chi_p$ is the nonequilibrium part.
We obtain for the collision integral (\ref{coll})
\begin{eqnarray}
\label{col}
{\rm St}f_p= {\rm St}f_0 -
\tau^{-1} \left(\chi_p - \frac{\langle
\chi_p \rangle }{\langle 1 \rangle}\right)
\frac{\partial f_0}
{\partial \varepsilon},
\end{eqnarray}
where the scattering rate
$$\tau^{-1} = T_{\ell}\sum_n
\langle  w^{(n)}_{pk} /\omega_k^{(n)}\rangle \sim \pi g^2 T_{\ell}.$$
The latter estimate is valid when
the ion temperature $T_{\ell}$ is considerably higher
than the Debye temperature;
the dimensionless electron-phonon coupling constant
$g \sim \lambda/\varepsilon_F \sim 1$. The
brackets denote  integration over the Fermi surface
$$\langle  ...\rangle  =\int \frac{2dS_F}{v(2\pi)^3}(...).$$
The first term in (\ref{col}) comes from the contribution
of the local equilibrium partition function:
\begin{eqnarray}
\label{coll0}
{\rm St} f_0 =
\sum_n \int
\frac{d^3p'}{(2\pi)^3} w_{pk}^{(n)}
\left(f_0(\varepsilon_p)-f_0(\varepsilon_{p'})\right)
\left( N_k^{(n)}(T_e) -N_k^{(n)}(T_{\ell}) \right)\nonumber\\
\times
\left( \delta(\varepsilon_p-\varepsilon_{p'} -\omega_k^{(n)})
+ \delta
(\varepsilon_{p}-\varepsilon_{p'}
+\omega_k^{(n)}) \right) .
\end{eqnarray}
This term describes the energy flow from electrons to phonons
when they are at different temperatures.
This term is absent if the temperatures of the electron and
lattice subsystems coincide.

\par The nonequilibrium part of the electron distribution function
has to  satisfy two conditions. The first  is indeed the
conservation law of the number of carriers:
$$\int\frac{d^3p}{(2\pi)^3}\chi_p\frac{\partial f_0}
{\partial \varepsilon}=0.$$
This expression determines the chemical potential and results
in the renormalization of the deformation potential:
$\lambda({\bf p}) \rightarrow \lambda({\bf p})- \langle
\lambda({\bf p}) \rangle /\langle 1 \rangle $.
\par The second condition
\begin{equation}
\label{norm}
\int\frac{d^3p}{(2\pi)^3}(\varepsilon_p-\mu)\chi_p\frac{\partial f_0}
{\partial \varepsilon}=0
\end{equation}
enables us to define the local temperature $T_e$ (see Ref. \cite{LLX}), i.e.,
to write the equation of thermoconductivity.

\par Substituting  Eq. (\ref{loc}) into the
Boltzmann equation, we get
\begin{equation}
\label{kin}
\frac{\partial \chi_p}{\partial t}+
{\bf v}\frac{\partial \chi_p}{\partial {\bf r}}+
\frac{\chi_p-
\langle \chi_p \rangle /\langle 1 \rangle}{\tau} = - e{\bf vE}-
\lambda_{ij}({\bf p})
\frac{\partial u_{ij}}{\partial t}+
\frac{\varepsilon_p-\mu}{T_e}\left(\frac{\partial T_e}{\partial t}+
{\bf v}\frac{\partial T_e}{\partial {\bf r}} \right) +
{\rm St}f_0
\left/ \frac{\partial f_0}{\partial \varepsilon} .\right.
\end{equation}
\par To obtain the equation for the local temperature $T_e({\bf r},t)$,
we
multiply the Boltzmann equation (\ref{kin}) by $(\varepsilon_p-\mu)
\partial f_0/\partial \varepsilon$ and integrate over ${\bf p}$.
With the help
of Eq. (\ref{norm}) we find the equation of thermoconductivity,
\begin{equation}
\label{tep}
c_e(T_e)\frac{\partial T_e}{\partial t}+{\rm div}\, {\bf q}
=Q-\alpha (T_e-T_{\ell}),
\end{equation}
where $c_e(T_e)=\pi^2 \langle 1 \rangle T_e/3
\equiv \beta T_e$ is the electron heat capacity
and ${\bf q}$ is the heat flow:
\begin{equation} \label{flow}
{\bf q}=\int\frac{2d^3p}{(2\pi)^3}{\bf v}(\varepsilon_p-\mu)
\chi_p \frac{\partial f_0}
{\partial \varepsilon}.
\end{equation}
Using Eq. (\ref{coll0}), we can find the last (relaxation)
term on the right-hand side of Eq. (\ref{tep}). For high temperatures
$T_{\ell}, T_e \gg \omega_D$, the electron--lattice relaxation
constant $\alpha$ is
$$
\alpha =\sum_n\int \frac{2dS_F dS_F'}{vv'(2\pi )^6}w_{pk}^{(n)}
\omega _k^{(n)}.
$$
\par
The density $Q$ of laser energy absorbed by electrons can
be taken in the form
\begin{equation}
\label{heat}
Q(z,t)=I(t)(1-R)\kappa e^{-\kappa z},
\end{equation}
where $R$ is the reflection coefficient and $\kappa$ is the inverse
penetration depth. The function $I(t)$ describes the pulse shape.
\par Eqiations (\ref{elas}) and ({\ref{tep})
must be supplemented by the proper boundary conditions.
We assume the simplest
geometry:  the metal
occupyies the half-space $z< 0$. Hence, the boundary conditions for the
above equations are
\begin{equation}
\label{boun}
\left.\frac{\partial T_e}{\partial z}\right|_{z=0}=0, ~~~~~
\left.\frac{\partial u_z}{\partial z}\right|_{z=0}=0,
\end{equation}
signifying that the heat flow through the surface and the normal stress
tensor component vanish on the surface.
\par We  also need the boundary conditions for the kinetic
equation (\ref{kin}) at the metal surface. These boundary conditions
depend on the type of   electron reflection  from  the surface.
We assume the specular reflection for simplicity.

\section{Dynamics of electron temperature and lattice deformation}

The above equations are nonlinear and very complicated. However, it is
possible to solve them in an important limiting case. Below we are
interested in times shorter than the electron--lattice relaxation time
$\tau_{e-l}\sim c_e(T_e)/\alpha$. In this case the lattice temperature
can be set  to the initial temperature $T_0$, and the last terms in Eqs.
(\ref{kin}) and (\ref{tep}) can be omitted.
\par To solve the system (\ref{elas}), (\ref{kin}) for the half-space
with the boundary
condition (\ref{boun}), we use the even continuation of
the temperature $T_e({\bf r}, t)$ and the partition function $\chi_p$,
and the odd continuation of $u_z({\bf r},t)$, into the half-space
$z< 0$:
\begin{equation}
\label{cont}
T_e(z<0)=T_e(-z,0),~~~
u_z(z<0)=-u_z(-z,0).
\end{equation}
For the parallel components $u_x$ and $u_y$ one must use the even
continuation, but owing to the fact that the external heat
(\ref{heat}) depends only on $z$, these components are vanish.
In Eq. (\ref{kin}) we discard $\langle \chi_p \rangle ,$
which represents the 'in-term' in the collision integral.
This term accounts for  carrier conservation, i.e., for
the isotropic channel of collision processes. Therefore it does not
affect the heat flow and lattice driving force.
\par The solution of Eq. (\ref{kin}) has the form
\begin{equation}
\label{solu}
\chi_p({\bf r},t)= \int\limits_{-\infty}^{t} dt'
X_p\left({\bf r}-{\bf v}(t-t'), t'\right)
\mbox{exp}\left({\displaystyle{-\frac{t-t'}{\tau}}}\right),
\end{equation}
where $X_p$ is the right-hand side of Eq. (\ref{kin}),
\begin{equation}
\label{tion}
X_p=-\lambda_{ij}({\bf p})
\frac{\partial u_{ij}}{\partial t}+
\frac{\varepsilon_p-\mu}{T_e}\left(\frac{\partial T_e}{\partial t}+
{\bf v}\frac{\partial T_e}{\partial {\bf r}} \right).
\end{equation}
Substituting the solution (\ref{solu}), (\ref{tion}) into the heat flow
(\ref{flow}) and integrating over the energy variable
according to $d^3p=d(\varepsilon_p-\mu)dS_F/v$, we obtain
\begin{equation}
\label{flow1}
{\bf q}({\bf r},t)=-\frac{\pi^2}{6}\left<  \int\limits_{-\infty}^{t}
dt'\exp\left({\displaystyle{-\frac{t-t'}{\tau}}}
\right){\bf v} \left(\frac{\partial}{\partial t'}
+{\bf v}\frac{\partial}{\partial {\bf r}}\right)
T_e^2({\bf r}-{\bf v}(t-t'), t')\right>.
\end{equation}
The expression (\ref{flow1}) is linear in $T_e^2$.
It is convenient to introduce the new function
$\Theta({\bf r},t)=T_e^2({\bf r},t)$  and  take the Fourier transform
with respect to space and time variables.
Then Eq. (\ref{flow1}) yields the Fourier component of the heat flow:
\begin{equation}
\label{flow2}
{\bf q}({\bf k},\omega) = \frac{i\pi^2}{6}\left<  \frac{(\omega-{\bf vk}){\bf v}}
{\omega-{\bf vk}+i\tau^{-1}}\right> \Theta ({\bf k},\omega).
\end{equation}
Substituting this result into the equation of thermoconductivity (\ref{tep}),
we obtain its Fourier component
\begin{equation}
\label{tepf}
-i\frac{\pi^2}{3}\left<\omega   +   \frac{(\omega-{\bf vk}){\bf vk}}
{\omega-{\bf vk}+i\tau^{-1}}  \right> \Theta ({\bf k},\omega)
=2\kappa (1-R) I(\omega) U({\bf k}),
\end{equation}
where $I(\omega)$ is the Fourier transform of the pulse shape $I(t)$.
The factor $Q({\bf k})$ describes the spatial distribution of the laser field
(\ref{heat}) ,
$$U(k_z) = \frac{2\kappa}{\kappa^2+k_z^2}$$
which depends only on $k_z$.
Equation (\ref{tepf}) yields the temperature dynamics
of metals under laser heating
with the time and
space dispersion.
\par We now turn to the equation for lattice
displacements (\ref{elas}).
The driving force $G_i({\bf r},t)$
can be evaluated as in  the  derivation of Eqs.
(\ref{flow1}) and (\ref{flow2}). Both the local equilibrium
partition function and nonequilibrium part (\ref{loc}) contribute to
the integral
(\ref{force1}). Expanding the integrals over the
energy variable in  powers  of $T_e/\varepsilon_F$ up
to the second order, we  obtain
\begin{equation}
\label{force}
G_i({\bf k},\omega) = -\frac{\pi^2}{6}\frac{\partial}{\partial
\varepsilon_F}\left<  \frac{\tau^{-1} \lambda_{ij}({\bf p})k_j}
{\omega-{\bf vk}+i\tau^{-1}}\right> \Theta ({\bf k},\omega).
\end{equation}
In addition to the electron force (\ref{force}) we also obtain  the
temperature-dependent renormalization of the elastic constants
$\lambda_{ijlm}$ (sound velocities) due to the interaction
with electrons (electron loop in the phonon self-energy function).
The dominant contribution in the range of interest comes from
the local equilibrium partition function:
$$
\lambda_{ijlm} \rightarrow \lambda_{ijlm} -\langle \lambda_{ij}({\bf p})
\lambda_{lm}({\bf p}) \rangle -\frac{\pi T_e^2}{6}
\frac{\partial^2}{\partial
\varepsilon_F^{2}} \langle \lambda_{ij}({\bf p})\lambda_{lm}
({\bf p}) \rangle .
$$
The electron contribution to the
sound velocity is  second
order in the electron temperature, $\Delta s/s \sim T_e^2/\varepsilon_F^2$.
\par
Taking the Fourier transform of the
left-hand side of Eq. (\ref{elas}), one needs to keep in mind
the singularity  at $z=0$ after  continuation (\ref{cont})
of the function $u_z$. This singularity
contributes the $d\delta(z)/dz$-term in the second derivative
$d^2u_z/dz^2$; such a term accounts for surface effects.
The Fourier transform with respect to
the coordinate  $z$ over the entire space gives
\begin{equation}
\label{elaf}
-\rho(\omega^2-s^2k^2)u_z (k,\omega)=G_z(k,\omega)+  k C(\omega),
\end{equation}
where $s=\lambda_{zzzz}/\rho$ is the longitudinal sound velocity
in the $z$-direction. In the last term, $C(\omega)$ must be
determined from the boundary condition
(\ref{boun}), and takes the form
\begin{equation}
\label{c}
C(\omega)=-2i\omega s \int \frac{dk}{2\pi}
\frac{G_z(k,\omega)}{\omega^{2}-s^2 k^2}.
\end{equation}
\par We next proceed to  the electron temperature and lattice deformations
represented by Eqs. (\ref{tepf}) and (\ref{elaf}) in
various limiting cases.

\section{Spatial variation of electron temperature and lattice deformation}

Equation (\ref{tepf}) describes the electron temperature
evolution  under ultrashort laser heating of
metals. This equation
generalizes the usual thermoconductivity equation \cite{ABE}.
We are interested in the wave vector $k$, which is
the greater of the inverse skin depth $\kappa$ ($\sim 10^5$ cm$^{-1}$)
and the electron
diffusion length $v \sqrt{\tau t_0}$ during the laser pulse  $t_0$.
In the usual experimental situation
$\tau^{-1}\sim 10^{14}$ s$^{-1}$, $\kappa v \sim 10^{13}$ s$^{-1}$,
and the hydrodynamic regime $\kappa v \ll \tau^{-1}$ obtains.
Thus, one can omit the term ${\bf kv}$ in the denominator
of the left-hand side of Eq. (\ref{tepf}). The
dominant contribution comes from the diffusion pole $\omega \sim \tau v^2k^2
\ll \tau^{-1}$. Therefore, we can also omit $\omega$ everywhere
in comparison with $\tau^{-1}$ or $kv$.
The solution of the thermoconductivity equation reads
\begin{equation}
\label{tem0}
\Theta(z,t)=\Theta_0+\frac{i}{\beta}
\int\limits_{-\infty}^{\infty}
\frac{dk\, d\omega \,dt'\,dz' }{(2\pi)^2 (\omega+iDk^2)} I(t')
e^{-i\omega(t-t')+ik(z-z') -\kappa \vert z' \vert},
\end{equation}
where the diffusion coefficient $D=\tau \langle {v_z^2}\rangle/\langle 1
\rangle$ is introduced.
The constant $\Theta_0=T_0^2$ comes from the solution of the corresponding
homogeneous equation, and represents the initial temperature.
Evaluating the integral (\ref{tem0}) with respect to $\omega$ and $k$,
we obtain
\begin{equation}
\label{tem}
\Theta(z,t)=\Theta_0+\int\limits_{-\infty}^{t}dt'\int\limits_{-\infty}^{\infty}
dz'\frac{Q(\vert z'\vert ,t')}{\beta \sqrt{\pi (t-t')D}}\mbox{exp}
\left( -\frac{(z-z')^2}{4(t-t')D}\right).
\end{equation}
We  see immediately that the function (\ref{tem}) satisfies
the boundary condition (\ref{boun}).
For the temperature at the surface $z=0$, Eq. (\ref{tem}) gives
$$T_e^{2}(0,t)=T_0^{2}+\frac{4}{\pi \beta}\int\limits_0^{t}dt'\,
Q(0,t-t')\,e^{\kappa^2 Dt'}\mbox{erfc}\left(\sqrt{\kappa^2 D t'}\right).$$
\par
The electron temperature (\ref{tem})
just after the pulse peaks at the surface:
\begin{equation}
\label{T}
T^2_{max} \sim \frac{It_0 (1-R)}{\beta} \mbox{min} \left(
\kappa, (Dt_0)^{-1/2} \right).
\end{equation}
This result has a simple explanation. For short pulses
$\kappa \sqrt{Dt_0}\ll 1$ the time dependence of the temperature
corresponds to the local laser intensity
 at the  observation point.
In the opposite case, $\kappa \sqrt{Dt_0}\gg 1$,
the temperature distribution  is determined mainly
by the diffusion process.

\par Consider now the equation for lattice displacements
(\ref{elaf}) with the force (\ref{force}).
Note that in the hydrodynamic regime, $\kappa v\ll\tau^{-1}$,
the dominant contribution to the
force $G_i$ comes from the
local equilibrium partition
function, i.e., the first term in (\ref{loc}),
if we consider times greater than the electron--electron
relaxation time, $t\gg\tau$.
In this case, the force has the simple expression
$$G_i({\bf r},t)= \Lambda_{ij} ~ \frac{\partial T^2_e({\bf r},t)}
{\partial x_j},$$
where the constants
$$
\Lambda_{ij}= \frac{1}{32\pi}\frac{\partial}{\partial \varepsilon_F}
\int\frac{dS}{v}\lambda_{ij}({\bf p}) \sim g\beta
$$
are of the order of the electron density of states at the Fermi surface.
\par From Eq. (\ref{elaf}) with the help of the expression (\ref{c})
one can find the lattice deformation
\begin{equation}
\label{def}
\frac{du_z}{dz} =\frac{i\Lambda_{zz}\kappa (1-R)
}{\rho \beta}\int\frac{d\omega dk}{(2\pi)^2}
\frac{k^2 U(k) I(\omega)}{(\omega+ik^2D)(\omega^2-s^2 k^2)}
\left[ e^{ikz} - e^{i\omega \vert z\vert/s}\right] e^{-i\omega t}.$$
\end{equation}
The first term in the brackets in Eq. (\ref{def}) represents
the particular solution of the inhomogeneous Eq. (\ref{elas})
while the second
corresponds to the general solution of
the homogeneous form of Eq. (\ref{elas}), and
represents the effect of the surface.
The integrand in (\ref{def}) contains  poles associated with the
diffuson and sound-wave excitations. Sound singularities
are bypassed using
infinitesimal phonon damping,  $\omega \rightarrow \omega+i0$.

\section{Effect of acoustic and optical displacements on destruction of
metals}

Equation (\ref{def}) describes the effect of nonequilibrium electron heating
on lattice deformations of acoustic type.
This deformation  vanishes at the
surface $z=0$ according to the boundary condition
(\ref{boun}).
For $z\ne 0$, the second term in brackets in (\ref{def}) represents
a deformation wave propagating from the surface into the bulk
of the  metal. It makes a nonzero contribution only at
sufficiently small depths $z <   st \sim 10^{-7}$cm.
Thus, we see that the deformation (\ref{def}) peaks
at $z\sim 10^{-7}$cm $\ll  \kappa^{-1}$.
To obtain the order of the effect, we can drop the second
term in  parentheses. It is then convenient to  integrate
over $\omega$, substituting the Fourier
transform $I(\omega)$.
We obtain
\begin{equation}
\label{defes}
\frac{du_z}{dz} \sim \frac{\Lambda_{zz}\kappa (1-R)
}{\rho \beta}\int\limits_0^t dt'I(t')
\int\frac{dk}{2\pi}
U(k)\left(\frac{e^{-i sk(t-t')}}{s(s+ikD)}-\frac{e^{-k^2D(t-t')}}{k^2D^2+s^2}
\right) e^{ikz}.
\end{equation}
Consider times greater than the duration of a pulse
$t> t_0$
but less than the characteristic time
of electron diffusion ($t< (\kappa^2 D)^{-1}\sim 10^{-12}s$)
and a sound-wave period
($t \ll (s \kappa)^{-1} \sim 10^{-11}s $)
with characteristic
wave vector of the order of the inverse skin depth $\kappa$.
In this range we can expand the exponentials in (\ref{defes})
in powers of $t$
up to   second order:
\begin{equation}
\frac{d u_z}{dz} \sim \frac{
\Lambda I t_0 \kappa (1-R)}
{2\rho\beta} t^2
\int\limits_{-\infty}^{\infty}\frac{dk}{2\pi}
U(k) k^2 e^{ikz}.
\end{equation}

Using the estimate $\Lambda/\rho \sim gs^2/\varepsilon^2_F$ and
the Eq. (\ref{T}) for  the maximum  electron temperature, we
extrapolate our result up to electron diffusion times
$t \sim (\kappa^2 D)^{-1} :$
\begin{equation}
\label{est}
\frac{du_z}{dz} \sim
g(1-R)\frac{It_0}{\kappa\beta}
\left(\frac{s}{\tau v^2\varepsilon_F}\right)^2 \sim
g\left(\frac{sT_{max}}{\kappa\tau v^2
\varepsilon_F}\right)^2.
\end{equation}
Setting
$s/v\sim 10^{-2}$, $\kappa \sim 10^5$~cm$^{-1}$, we
arrive at the numerical estimate
$du_z/dz \sim 10^{-2} g^5 (T_e/\varepsilon_F)^2$.
\par
Our result contains the natural factor
$T^2/\varepsilon_F^2$,
which means that  laser heating is important as soon
as the electron temperature is higher than the Fermi energy.
Although the estimate was obtained for $T_e \ll \varepsilon_F$,
it is still roughly correct up to $T_e \sim\varepsilon_F$.
The additional small factor $s^2/(v\kappa l)^2$ is due to the
fact that the characteristic period of the sound wave ($10^{-11}s$) is
much greater than the charachteristic times of electron diffusion
($10^{-12}s$) and laser heating ($10^{-13}s$).
Therefore it would be of considerable interest to calculate the lattice
deformation from high-frequency (but  long-wavelength) excitations,
i.e., optical phonons whose period is about $10^{-14}s$.
This case differs from the calculations  above
in the equation of lattice motion  (1) and electron force (2),
due to the different form of deformation potential ( see \cite{MF} ).
Estimates show that the relative
optical displacement
(with respect to the lattice
constant)
are of order
$T^2/\varepsilon_F^2$.

\section{Conclusions}

\par Our result for acoustic deformation
(\ref{est}) agrees with the experiment reported by P. Rentzepis
[17] where a deformation  $du_z/dz \sim 10^{-3}$
had been observed in the laser heating of noble metals.
However, we see that the interaction of heated electrons with optical phonons
can provide a more effective means of strong lattice deformation,
but this case has yet to be studed experimentally.
An ultrashort intense laser pulse
can result in the
destruction and ablation of metals,
while only the electron component is heated,  and the
lattice stays  at considerably low temperature.
\par In conclusion, we  emphasize two
points. First, as follows from Eq. (9), the driving force
for  lattice expansion is proportional to
$T_e \partial T_e/\partial z$. Because of the high
absorption coefficient of metals in the UV
($\kappa\sim 10^5 $cm$^{-1}$), the temperature
gradient reaches $\sim 10^9$~K/cm. Note that
the extremely  high values of this parameter (which
 is typical
of metals) leads to nonequilibrium expansion
of the lattice.      Second,  subpicosecond
elastic deformation of the lattice of  order
$10^{-3}-10^{-2},$ corresponding to an internal pressure
$10-100$ GPa, can provide an effective mechanism for
subsequent laser fracture of metals.

\par We are grateful to S. I. Anisimov and V. A. Benderskii
for many useful discussions and valuable comments. This work was
supported in part by the Russian Foundation for Basic Research, Grant
N$_o$ 97-02-16044.
One of the authors (EGM) also thanks KFA Forschungszentrum, J\"{u}lich,
Germany, for a Landau Postdoctoral Fellowship.



\begin{thebibliography}
\noindent
\bibitem{ABE} S. I.  Anisimov, A. M.  Bonch-Bruevich, M. A.  El'yashevich
et al., Sov.  Phys. Tech.  Phys.  {\bf 11}, 945 (1967);
S. I.  Anisimov, B. L.  Kapeliovich, and T. L.  Perelman,
Zh. \'{E}ksp. Teor. Fiz. {\bf 66}, 776 (1974).
\bibitem{YLB} R.  Yen, J.  Liu, and N.  Bloembergen,
Opt.  Commun.  {\bf 35}, 277 (1980);
R.  Yen, J.  Liu, N.  Bloembergen et  al., Appl.  Phys.  Lett.  {\bf 40},
185 (1982).
\bibitem{KLT} M. I.  Kaganov, I. M.  Lifshits, and L. V.  Tanatarov,
Zn. \'{E}ksp. Teor. Fiz. {\bf 31}, 232 (1957).
\bibitem{EPN} H.  Elsayed-Ali, M.  Pessot, T.  Norris, and G.  Mourou,
in {\it Ultrafast Phenomena V}, Proc.  of the Fifth Optical Society
of America Meeting, Snowmass, Colorado, 1986,   G. R.  Fleming
and A. E.  Siegman (eds.), Springer Series in Chemical Physics, Vol. 46,
Springer-Verlag, New York (1986), p. 264.
\bibitem{FLI} I. G.  Fugimoto, J. M.  Liu, E. P.  Ippen et al., Phys.  Rev.
Lett. {\bf 53}, 1837 (1984).
\bibitem{RWD} D. M. Riffe, X. Y.  Wang, and M. C.  Downer, J.  Opt.  Soc.
Amer. B {\bf 10}, 1424 (1993).
\bibitem{GG} J. P.  Girardeau-Montaut and C.  Girardeau-Montaut, Phys.  Rev.
 B {\bf 51}, 13560 (1995).
\bibitem{ABF} S. I.  Anisimov, V. A.  Benderskii, and G.  Farkas,
Usp. Fiz. Nauk {\bf 122} (2) 185 (1977).
\bibitem{KKH}A. G.   Krivenko, J.  Kruger, W.  Hautex et al.,
Benchte Bunsengeselsch.  Phys.  Chem.  {\bf 99}, 1489 (1995).
\bibitem{B} V. A.  Benderskii, Russian Electrochem.  {\bf 33}, 417 (1997).
\bibitem{KKH'}A. G.  Krivenko, J.  Kruger, W.  Hautex et  al.,
Russian Electrochem.  {\bf 33}, 426 (1997).
\bibitem{ABG} M. B.  Agranat, A. A.  Benditskii, G. M.  Gandel'man  et al.,
JETP Lett.  {\bf 30}, 167 (1979).
\bibitem{AAM} M. B.  Agranat, S. I. Anisimov, and B. I.  Makshantsev, Appl.
Phys.  B {\bf 47}, 209 (1988).
\bibitem{AAM'} M. B.  Agranat, S. I. Anisimov, and B. I.  Makshantsev, Appl.  Phys.  B
{\bf 55}, 451 (1992).
\bibitem{AR} S. I.  Anisimov and B.  Rethfeld,  Proc. SPIE
 {\bf 3093}, 192 (1997).
\bibitem{FM} L. A.  Falkovsky and E. G.  Mishchenko, JETP Lett.
{\bf 66}, 195 (1997).
\bibitem{R} P.  Rentzepis, in Abstracts of the Second Conference
on Modern Trends in
Chemical Kinetics, Vol. 1, Novosibirsk, Russia (1995), p. 142.
\bibitem{FM'} L. A.  Falkovsky and E. G.  Mishchenko, Phys.  Rev.  B {\bf 51}, 7239 (1995).
\bibitem{K} V. M.  Kontorovich,
Usp. Fiz. Nauk {\bf 142} (2), 265 (1984).
\bibitem{LLX} E. M.  Lifshitz and L. P.  Pitaevski, {\it Physical Kinetics},
Nauka, Moscow (1979), p. 33.
\bibitem{MF} E. G. Mishchenko and L. A. Falkovsky,
Zh. \'{E}ksp. Teor. Fiz. {\bf 107}, 936 (1995).

\end{thebibliography}
\end{document}